\documentclass[onecolumn,11pt]{article}
\usepackage[english]{babel}
\usepackage[utf8x]{inputenc}
\usepackage[pdftex]{graphicx}
\usepackage[a4paper, top=1in, bottom=1.25in, left=0.8in, right=0.8in]{geometry}
\usepackage{enumitem}
\usepackage{amsmath}
\usepackage{amssymb}
\usepackage{cite}
\usepackage{bm}
\usepackage{braket}
\usepackage{float}
\usepackage{siunitx}
\addto{\captionsenglish}{%
	
}

\graphicspath{{./Figures/}}
\setlist{leftmargin=*}

\begin{document}

\title{\textbf{Long-range order in quadrupolar systems on spherical surfaces}}
\author{Andraž Gnidovec and Simon Čopar \\ \textit{University of Ljubljana, Faculty of Mathematics and PhysicsSI-1000 Ljubljana, Slovenia}}
\date{}

\maketitle

\begin{abstract}
Understanding the interplay between topology and ordering in systems on curved manifolds, governed by anisotropic interactions, takes a central role in many fields of physics. In this paper, we investigate the effects of lattice symmetry and local positional order on orientational ordering in systems of long-range interacting point quadrupoles on a sphere in the zero temperature limit. Locally triangular spherical lattices show long-range ordered quadrupolar configurations only for specific symmetric lattices as strong geometric frustration prevents general global ordering. Conversely, the ground states on Caspar-Klug lattices are more diverse, with many different symmetries depending on the position of quadrupoles within the fundamental domain. We also show that by constraining the quadrupole tilts with respect to the surface normal, which models interactions with the substrate, and by considering general quadrupole tensors, we can manipulate the ground state configuration symmetry.
\end{abstract}

\section{Introduction}

One of the central concepts in studying the electrostatic interactions between building blocks of matter is the multipolar expansion of electric and magnetic potentials. The significant terms in this expansion depend on the charge or polarization distribution, as well as the shape of interacting particles \cite{Vedmedenko2008}. Objects with high symmetry only have a small number of nonvanishing multipolar moments and the expansion can be truncated after the first significant term as subsequent contributions are often small in comparison. In many cases, however, higher multipole contributions can have determining effect on system behavior. They play an important role in nonlinear optical phenomena \cite{Shen2003}, interactions between molecules \cite{Buckingham1967,Birnbaum1985} and can significantly alter orientational configurations in 2D magnetic systems \cite{Vedmedenko2005,Ewerlin2013,Mikuszeit2009}.

An important class of systems interacting predominantly via multipole interactions are molecular adsorbates of H$_2$, N$_2$, O$_2$ or CO \cite{Berlinsky1978,OShea1979,OShea1982,Mouritsen1982,Pan1982,Peters1985,Kim1999,Sullivan2000,Boyd2002,Ustinov2016,Ustinov2018}, and molecular crystals \cite{Hamida1994,Raugei1997}, where quadrupole moments of positionally fixed building blocks dictate the orientational ordering in the system. The energetically preferred configuration of two quadrupoles is the mutually orthogonal T-configuration. This leads to high geometrical frustration in systems where the symmetries of interaction and lattice are not compatible, e.g. quadrupolar glasses \cite{Sullivan1978,Sullivan1988,Kim1997} and systems of quadrupoles on 2D triangular lattices. Such systems consequently show complex orientational ordering in low temperature states, as well as rich phase behavior. As found from theoretical considerations \cite{Berlinsky1978,Harris1984,Sullivan2000} and corroborated by simulations \cite{OShea1979,Klenin1982,Mederos1990}, the ground state of a quadrupolar system on a triangular lattice is a long-range ordered pinwheel structure. If additional interaction with the substrate is present that favors in-plane ordering of quadrupoles, the ground state assumes the herringbone structure instead \cite{Berlinsky1978,Massidda1984,Mederos1990,Marx1993}.

Whereas the behavior of systems with multipolar interparticle interactions is well understood in Euclidean geometries, in recent years there has been a general surge of interest in systems on curved surfaces governed by different anisotropic interactions. The sphere in particular is one of the prototype systems to study the effects of curvature on system behavior due to its unique topological properties \cite{Bowick2009}. Intriguing structural order supported by curvature and interaction anisotropy emerges in liquid crystals \cite{Vitelli2006, Li2013}, dynamical systems \cite{Sknepnek,Praetorius} and thin magnetic spherical shells \cite{Kravchuk2016,Milagre2007,Sloika2017}. Furthermore, orientational ordering in systems with pure multipolar interaction, specifically the dipole-dipole interaction, has also recently been studied on spherical lattices \cite{Gnidovec2020,Copar2020}. It was shown that the local positional order and the symmetry of the lattice play a fundamental role in determining the symmetries of the ground state and excited state configurations.

As spherical lattices lack any translational symmetries and also contain structural defects due to topological constraints, the pinwheel and herringbone quadrupole global ordering are in general not possible on a sphere even in cases where the lattice locally resembles the triangular lattice (e.g. the Thomson lattice). The absence of translational symmetry was previously shown to break long-range ordering in some 2D systems of quadrupoles~\cite{Vedmedenko2008_1}. Theoretical studies of systems on spherical lattices can therefore provide further insight on the effects of local order, symmetries, as well as curvature, on long-range quadrupole ordering. Furthermore, possible experimental realizations of quadrupolar systems on a sphere include CO$_2$ and ethylene adsorbates on spherical fullerenes, where quadrupole moment of molecules was also found to influence orientational order \cite{MartnezAlonso2000,Ralser2016,Zottl2014}. 

In this work, we use numerical modeling to examine stable configurations of point quadrupoles, positionally fixed to different spherical lattices. We first define the interaction potential and describe the general parametrization of quadrupole tensors along with simulation methods used to acquire stable configurations. We then present the results of ordering on individual Thomson lattices, showcasing the influence of lattice symmetry on long-range orientational order in quadrupolar systems. We compare the results to that on the icosahedral Caspar-Klug (CK) lattices to further examine the effects of local positional order and geometrical frustration on quadrupolar ordering. Finally, a variant of the problem with constrained tilt between the quadrupoles and the sphere is considered as it more closely resembles some of the possible experimental realizations.

\section{Simulation methods}

We consider a system of $N$ point quadrupoles, described by Cartesian tensors $\{Q^A\}$ and positionally fixed to lattice points $\{\bm{r}_A\}$. The interaction energy functional of the system in units of $1/4\pi\epsilon_0$ can be written as a quadratic form,
\begin{equation}\label{energy}
\mathcal{U} = \frac{1}{2}\sum_{A\neq B} U_{ijkl}(\bm{r}_{AB}) Q^A_{ij} Q^B_{kl},
\end{equation}
where $\bm{r}_{AB} = \bm{r}_B - \bm{r}_A$ and Einstein summation is implied for lowercase indices. We consider interactions between all pairs of quadrupoles. The interaction tensor
\begin{equation}
U_{ijkl}(\bm{r}) = \frac{1}{3r^5} \left[35\frac{x_i x_j x_k x_l}{r^4} - 20 \frac{x_j x_k \delta_{il}}{r^2} + 2 \delta_{ik}\delta_{jl}\right]
\end{equation}
can be computed in advance as the positions of quadrupoles are fixed which can significantly decrease energy calculation times. Note that the interaction in consideration is purely classical which provides a sufficient description of orientational ordering in most quadrupolar molecular adsorbates \cite{Vedmedenko2007}.

Most commonly, quadrupole moments found in nature are of the rotationally symmetric linear (uniaxial) type with eigenvalues $e_1=-2e_2=-2e_3=1$. Unless specifically mentioned, this is the type we focus on throughout this paper. For a complete description, we later also consider a continuous parametrization of all possible quadrupole eigentensors, with eigenvalues $e_1=2+a$, $e_2=-1-2a$ and $e_3=-1+a$ where $a\in[0,1]$. The case of $a=0$ corresponds to the linear type whereas $a=1$ describes planar quadrupoles with eigenvalues $e_1=-e_2=1$ and $e_3=0$ which are not extensively studied in literature. In energy calculations, we additionally rescale the parameterized eigenvalues so that $\text{Tr}(Q^2) = 1$ which can be formally done by introducing a constraint to minimization of Eq.~(\ref{energy}).

We parameterize the orientations of uniaxial quadrupoles in their respective local coordinate frames by two angles, azimuthal angle $\phi_A$ and polar angle $\theta_A$. However, this parametrization is not sufficient to describe orientations of biaxial quadrupoles where the description with Euler angles is used instead. We use numerical minimization of energy with respect to the parametrization angles to find the ground state configurations of the system. Our algorithm of choice is the Broyden-Fletcher-Goldfarb-Shanno algorithm which proved effective as well as time efficient in finding the ground state and excited state configurations for smaller system sizes ($N < 150$). Although previous studies of quadrupolar ordering mostly relied on the simulated annealing Monte Carlo simulations to find the ground state structure of the system \cite{Vedmedenko2008, Vedmedenko2008_1}, the use of a deterministic minimization method allows us to more systematically examine some excited state configurations. We performed 1000 minimization simulations at each $N$, starting from different random initial conditions to ensure finding the global minimum as well as to find excited states with distinct symmetries. Nevertheless, finding the ground state in this way becomes increasingly difficult with the increase in system size as the number of metastable states grows exponentially with $N$. The limit $N=150$ represents the approximate size above which the deterministic minimization becomes sub-optimal. Difficulties in finding a global minimum numerically arise due to the rough energy landscape. This would also reflect in a physical system where we would most likely find in a glassy, disordered state instead of the true ground state.

Finding the true ground state from random initial orientations of quadrupoles can be inconsistent on some non-uniform lattices that deviate strongly from equilateral positional order, due to the fast decay of the interaction strength with the distance that promotes metastable states on such lattices. In particular, this problem emerges for many CK lattices. To circumvent it, we additionally performed minimization simulations with constrained orientational order symmetry, considering all symmetry subgroups of the icosahedral group of the underlying lattice which drastically improves the quality and consistency of the results. Such constrained minimization is also computationally more efficient as imposing symmetries to the resulting orientational order decreases the number of degrees of freedom in the system.

Due to the complex ordering of quadrupolar systems on the sphere, no single order parameter can itself sufficiently quantify the long range order. As we will demonstrate in the following sections, some properties of the quadrupolar systems can be deduced based on the out-of-plane parameter
\begin{equation}\label{out-of-plane}
\xi = \frac{1}{N}\sum_{i=1}^N \xi_i = \frac{1}{N}\sum_{i=1}^N |\hat{\bm{k}}_i \cdot \hat{\bm{r}}_i|,
\end{equation}
where $\hat{\bm{k}}$ is the direction of the main quadrupolar axis, evaluating the average of the deviation $\xi_i$ of quadrupole orientations from tangential ordering. A more comprehensive quantitative approach to differentiate between quadrupolar states with different degrees of ordering is the expansion of their orientations over tensor spherical harmonics \cite{Thorne1980}. This expansion can be defined analogously to the definitions of scalar and vector spherical structure factors used for investigating scalar and vector order on the surface of the sphere \cite{Franzini2018,Bozic2019,Copar2020}. 
However, comparison between configurations at different $N$ still proves to be challenging because of relatively small system sizes that strongly influence expansion coefficients, especially at higher orders in the expansion. As will be shown in the following sections, long-range order in quadrupolar systems is adequately described by state symmetries and, therefore, tensor spherical harmonic analysis remains beyond the scope of this article.

\section{Results}

\begin{figure}[!ht]
	\centering
	\includegraphics[width=0.5\linewidth]{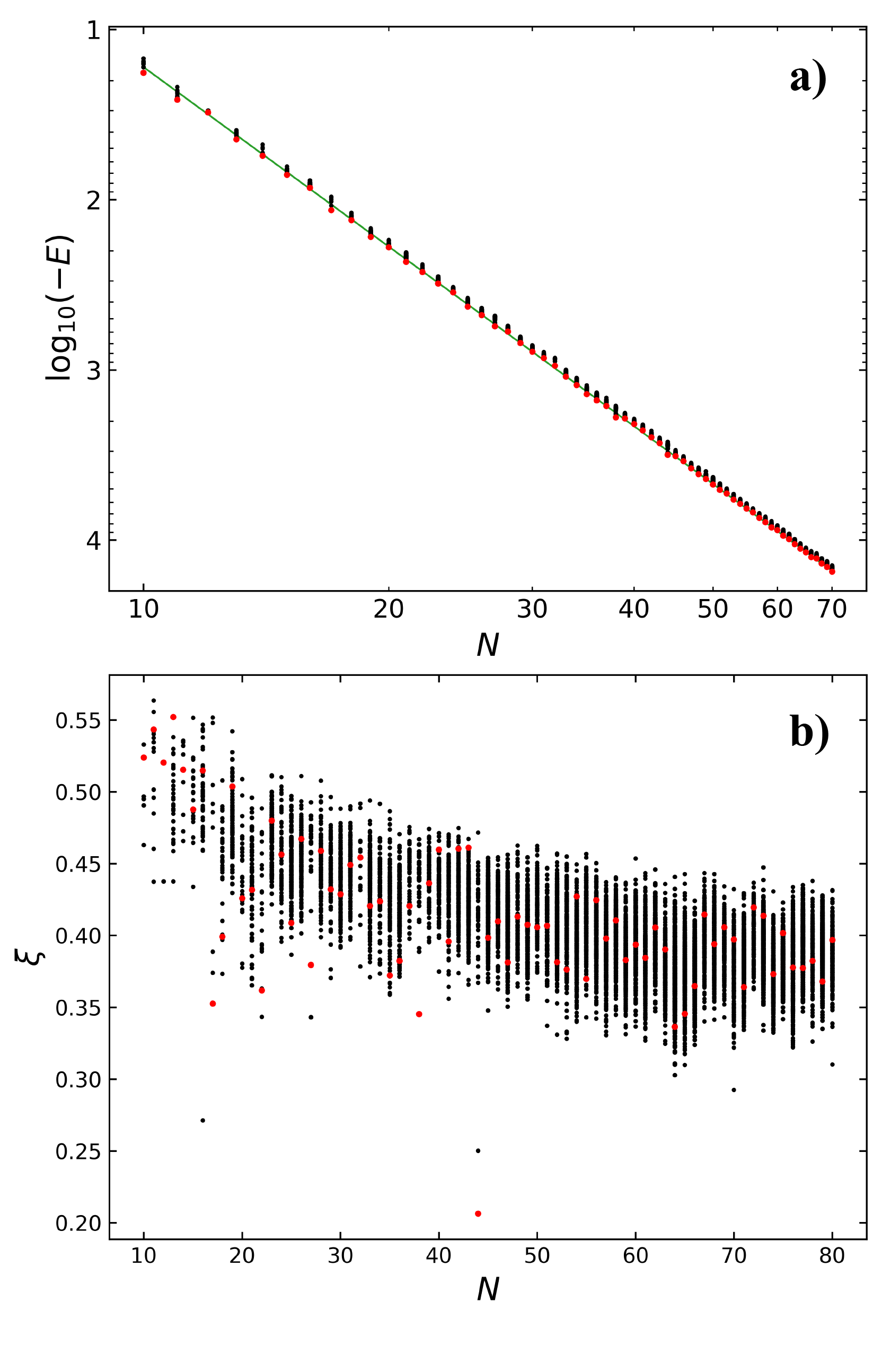}
	\caption{(a) Energies and (b) the deviation from tangential ordering $\xi$ for quadrupolar states on the Thomson lattices from $N=10$ to $N=78$. Red dots represent ground state values whereas black dots correspond to excited state configurations. Ground state energies follow a power law trend $E_0\propto N^{3.48}$, presented in the log-log plot as a blue straight line. Ground state values of $\xi$ vary significantly with the system size, however, the average value of $\xi$ over all configurations at each $N$ is decreasing. The outliers at $N=16$ and $N=44$ with the lowest $\xi$ correspond to pinwheel configurations.}
	\label{fig:en_xi}
\end{figure}

We study the orientational ordering of quadrupoles, positionally fixed to different spherical lattices, and explore the general properties and possible symmetries of the ground state and excited state configurations. First, we focus on uniform lattices, in particular the Thomson lattice, taken as a representative locally triangular case with many distinct symmetries at different system sizes which proves to be fundamental for high symmetry quadrupolar ordering. Figure~\ref{fig:en_xi} shows the dependence of the ground state energies on the size of the underlying Thomson lattice $N$ that determines the positional order in the system. Fitting the power law curve shows scaling as $E_0\propto N^{3.48}$, closely matching the exponent estimate of $7/2$ calculated from the neighbor distance and energy scaling $r\propto N^{-1/2}$ and $E\propto N/r^5$, respectively. Another quantity showing distinct trend with increasing system size is the out-of-plane parameter $\xi$ (Eq.~\ref{out-of-plane}).
Whereas the ground state values, marked by red dots, vary significantly with $N$, the average values of $\xi$ over all configurations at each $N$ are smaller for bigger systems. Note that in the 2D pinwheel configuration $\xi=1/4$\cite{} which already shows that the ordering in spherical case is more complex. Additionally, the decrease in $\xi$ with $N$ is consistent with the assumption that in the limit of large system sizes, the role of curvature on ordering is small and the pinwheel structure again becomes the ground state of the system, at least in regions away from disclinations and dislocations. 

Large deviations in ground state values of the out-of-plane parameter $\xi$ indicate that there is no general structural order in quadrupolar systems on the Thomson lattice which can be quickly confirmed by examining individual ground state configurations. This is in stark contrast to the dipolar case where a general macrovortex configuration was found (cf. \cite{Gnidovec2020}) but it is not surprising as dipolar configurations are not geometrically frustrated on locally triangular lattices. Nevertheless, we find that many individual configurations with symmetric quadrupolar order are possible on certain lattices with high positional order symmetry, not only in the ground state but also in excited state configurations.

\begin{figure}[!ht]
	\centering
	\includegraphics[width=0.7\linewidth]{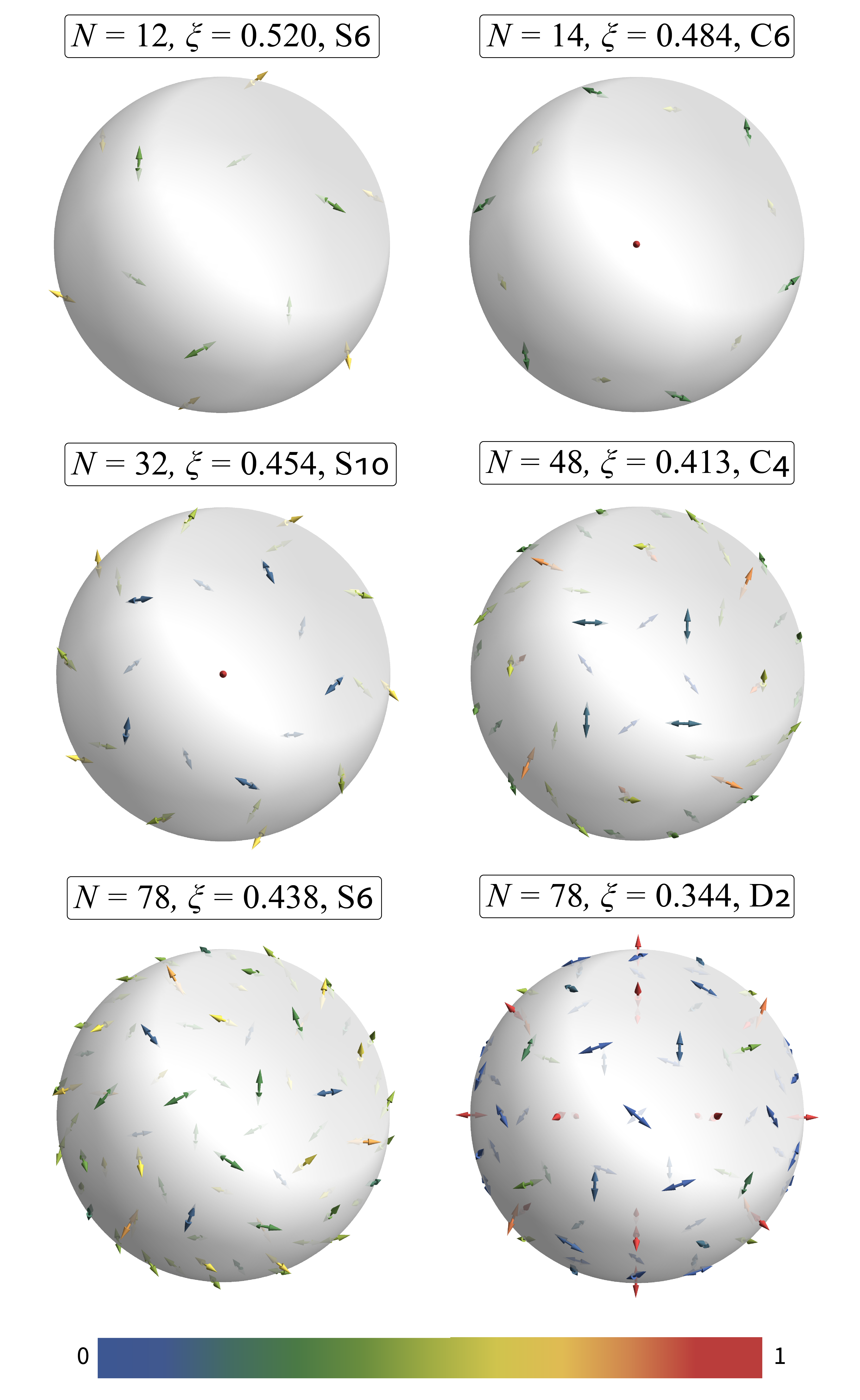}
	\caption{Examples of rotationally symmetric configurations on the Thomson lattices: the ground state for $N=12$ ($S_6$), one of the excited states for $N=14$ (C$_6$), the ground states for $N=32$ (S$_{10}$) and $N=48$ (C$_4$), and two excited states of $N=78$ (S$_6$ and D$_2$). The quadrupoles are represented by double arrows and colored according to the projection of its orientation to the surface normal $\xi_i$. Radial quadrupoles are colored in red and tangential in blue.}
	\label{fig:quad_rot}
\end{figure}

Some of these structures are presented in Fig.~\ref{fig:quad_rot} where quadrupoles are colored depending on the inclinations with their respective tangent planes, $\xi_i$. More information on ordering at different system sizes is available in Supplemental Material\cite{SuppMat} (SM). We first take a look at the ground state configuration for $N=12$ with icosahedral positional order. Quadrupoles order in a quasi-vortex configuration with three-fold rotational and inversion symmetries ($S_6$ point group), with all quadrupoles showing considerable out-of-plane ordering. This results in the highest value of $\xi$ among all ground state configurations found, at $\xi=0.520$. Similar vortex ordering is possible for $N=14$, where it actually resembles the pinwheel structure with quadrupoles on the symmetry axis oriented radially to the surface. However, this $C_6$ symmetric configuration is not the ground state. In fact, all 8 other configurations found have lower energies and are less symmetric; the ground state has no symmetries and three other excited states show $C_2$ symmetry (symmetry axis is the same as for the $C_6$ case).

Increasing the system size, we find many other cases of axial symmetries both for the ground state and excited state configurations. To illustrate this, we show the ground state configurations belonging to $C_4$ point group found on the octahedral $N=48$ lattice and $S_{10}$ point group on the icosahedral $N=32$ lattice. The latter includes two pinwheels centered at the poles of the sphere. Note that some excited states for the $N=48$ lattice show other symmetries as well, i.e. $D_4$, $C_2$ and $C_3$. We can find intriguing configurations also for the case of $N=78$ with tetrahedral positional symmetry. The quadrupolar ground state is not shown here as it is less interesting with only $C_2$ symmetry and excited states showing higher degrees of orientational ordering. In particular, the $S_6$ configuration is structurally similar to the ground state of $N=12$ with quasi-vortex ordering and the highest value of out-of-plane parameter at this system size, $\xi=0.438$. In contrast, we also find a configuration where ordering gravitates towards a pure pinwheel structure, though some quadrupoles still deviate from in-plane or radial orientations. This provides another demonstration that the long-range nature of quadrupolar interaction influences local ordering on curved surfaces. The configuration belongs to the D$_2$ point group, with three distinct 2-fold rotation axes that are perpendicular to one another, inherited from the tetrahedral group of the positional order.

The ground state for $N=32$ deserves additional discussion as it can be related to existing experimental and simulational investigations. Adsorbates of nonpolar molecules, such as H$_2$, N$_2$ and ethylene, can form a $1\times 1$ commensurate state as they adsorb onto fullerene C$_{60}$, where all pentagons and hexagons are occupied by exactly one molecule of the adsorbate \cite{Calvo2012,Echt2013,Kaiser2013,Zottl2014}. The energetically favored adsorption sites are the 32 centers of carbon polygons that form exactly the same icosahedral configuration as the $N=32$ Thomson solution. Orientational order of adsorbed ethylene molecules in such $1\times 1$ commensurate configuration was investigated in a molecular dynamics (MD) simulation in Ref.~\cite{Zottl2014}. It was found that at low temperature (\SI{1}{K}), two molecules orient perpendicular to the surface of the fullerene whereas other 30 lie almost tangentially. This result resembles the quadrupolar ground state configuration presented in Fig.~\ref{fig:quad_rot}, however, our result shows deviations from tangential ordering for quadrupoles around the equator. This discrepancy can be attributed to different causes. MD simulations also take into account the interaction with the substrate, and additionally, the ethylene molecule is not rotationally symmetric so modeling with linear quadrupoles does not provide a complete description of ordering. We will further comment on the solutions for different quadrupole tensors in Sec.~\ref{constr_sol}.

\begin{figure}[!ht]
	\centering
	\includegraphics[width=0.9\linewidth]{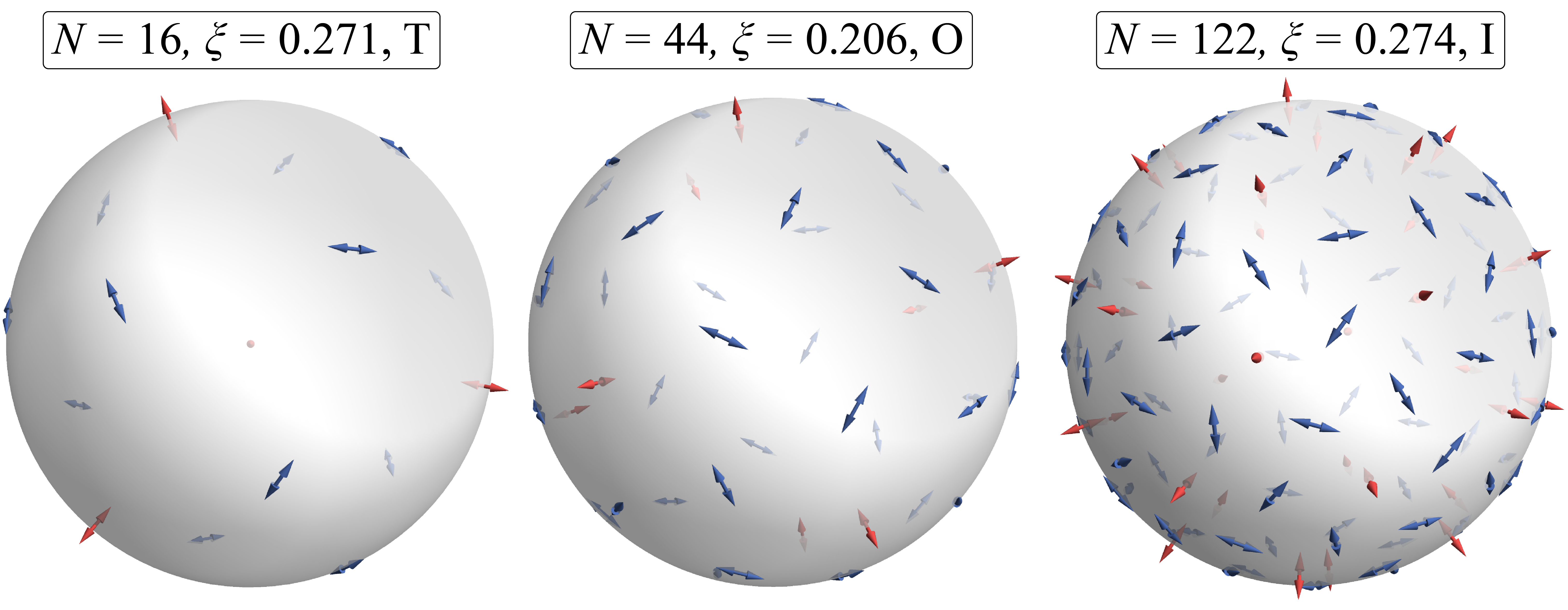}
	\caption{High symmetry configurations on the Thomson lattices: tetrahedral $N=16$, octahedral $N=44$ and icosahedral $N=122$. Whereas the last two are the ground states for the systems of their respective sizes, the tetrahedral configuration only emerges as an excited state for $N=16$.}
	\label{fig:quad_sym}
\end{figure}

The last presented configuration in Fig.~\ref{fig:quad_rot} indicates that pure pinwheel structures could be possible on certain Thomson lattices. This is indeed the case, in Fig.~\ref{fig:quad_sym} we present three such structures at different system sizes, $N=16$ (not the ground state), $N=44$ and $N=122$. What is more, these quadrupolar configurations preserve the symmetries of their positional order, tetrahedral, octahedral and icosahedral, respectively. It is immediately clear looking at this configurations that pure pinwheel ordering is not compatible with every Thomson lattice and emerges only for specific arrangements of lattice defects. The perpendicular quadrupoles should be centered at all lattice disclinations (as for $N=122$) or completely avoid them (e.g. for $N=16$ and $N=44$). The arrangement of defects determines if constructing a complete pinwheel lattice is possible with these conditions. It should be pointed out that quadrupoles in these pinwheels deviate slightly from the tangential orientations. Nevertheless, the out-of-plane parameter $\xi$ is significantly smaller for these configurations than the others at the same system size which is also reflected in Fig.~\ref{fig:en_xi}.

We showed that certain symmetric Thomson lattices support long-range ordered symmetric quadrupolar configurations, however, these should be searched for on per lattice basis. Examining stable configurations on the Tammes and Fibonacci lattices, we further confirm this observation (see SM). The Tammes lattice is similar to the Thomson lattice, i.e. it is locally composed of close-to-equilateral triangles. This characteristic was found to be central in determining the orientational order in discrete dipolar systems (cf. Ref.~\cite{Gnidovec2020}). In quadrupolar systems on a sphere, it leads to strong geometric frustration, especially considering the topological requirement for lattice defects. The ordering on the Tammes lattice is therefore expectedly similar to the results presented above, with only individual long-range ordered quadrupolar configurations, confirming that the exact positional order is less important than its global properties such as the symmetry. On the other hand, quadrupolar ordering changes if the lattice deviates systematically from local triangular order. An example is the Fibonacci lattice, where for high enough system sizes, it locally resembles the square lattice around the equator and, consequently, for $N \gtrsim 70$ the quadrupoles order in the local windmill configuration. The same ordering was found in the mean-field calculations for the 2D square lattice \cite{Massidda1984}. On the poles of the Fibonacci lattices where positional order is highly irregular, long-range quadrupolar order is again absent.

\subsection{Icosahedral Caspar-Klug lattices}

Investigation of stable states on the Thomson lattices showed that the symmetrically nontrivial orientational ordering in both the ground states and excited states is mainly connected to the symmetry of the positional lattice, at least for lattices that are locally composed of (close to) equilateral triangles. We further explore the role of local order and global positional arrangement on quadrupolar configurations by examining ordering on the CK lattices \cite{Bruinsma2015,Rochal2017} where we can manipulate local positional order while retaining the icosahedral lattice symmetry. We focus on CK lattices with the triangulation parameters $(n, m) = (1, 0)$, resulting in system size $N=60$, at different values of parameters $(u, v)$ that determine the positions of quadrupoles inside the fundamental domain (see Ref.~\cite{Copar2020} for more information on lattice construction). These lattices are intrinsically non-uniform, with vacancies in the positional order even in the equidistant cases (constant nearest neighbor distances) which leads to ordering that is markedly different than in the Thomson case studied above.

\begin{figure*}[!ht]
	\centering
	\includegraphics[width=\linewidth]{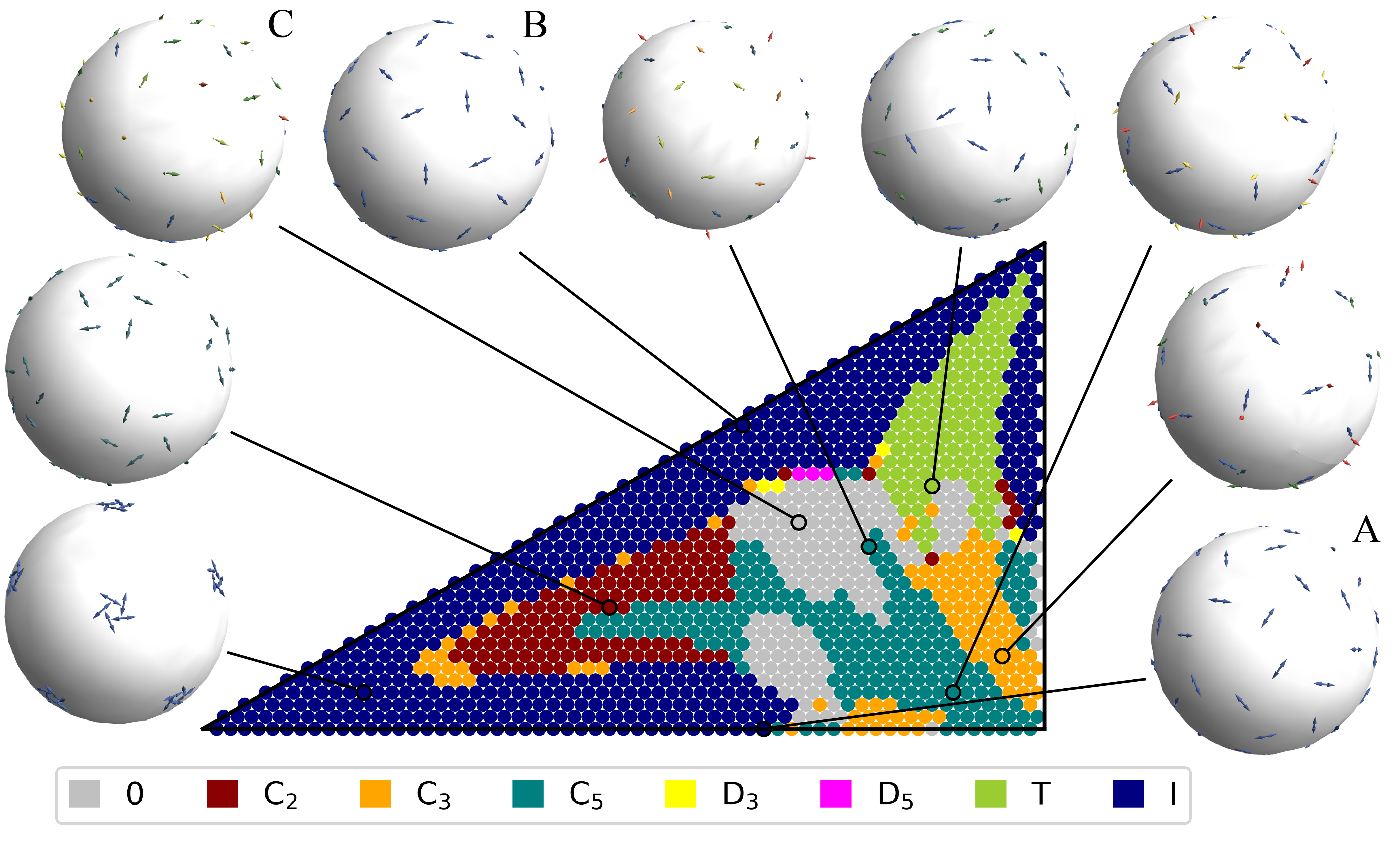}
	\caption{A diagram of ground state symmetries depending on lattice parameters $u$ and $v$ for 1239 points over the whole fundamental domain. Additionally, examples of quadrupolar configurations are also shown, including the equidistant lattices A, B and C, among others. Orientational order symmetries depend strongly on local positional order. Strong geometric frustration in the middle of the diagram prevents the formation of long range order, eg., the lattice C, whereas the configurations towards the edges show many different uniaxial and even higher symmetries, such as the icosahedral states on lattices A and B.}
	\label{fig:CK_symmetries}
\end{figure*}

Quadrupolar ground states on some equidistant CK lattices are presented in Fig.~\ref{fig:CK_symmetries}, marked as A (truncated icosahedron), B (small rhombicosidodecahedron) and C (snub dodecahedron). The lattice A supports an icosahedral quasi-pinwheel structure around each vertex (five-fold axis) of the icosahedron, with a missing perpendicular central quadrupole. Quadrupoles are oriented close to their tangential planes which results in $\xi\approx 0.1$, a value much lower than for any configuration on Thomson lattices. The configuration on the lattice B is similar to the lattice A, with only the arrangement of the quasi-pinwheels being different. The lattice C, however, does not support any symmetries in the ground state despite equal distances between quadrupoles. It consists of pentagons and equilateral triangles between them which leads to geometric frustration of quadrupole orientations and breaks global ordering.

We conducted symmetry analysis of the ground state configurations in dependence to the position of quadrupoles in the fundamental domain as determined by the parameters $u$ and $v$. To ensure finding the true ground state of the system, we additionally performed the simulations with constrained orientational order symmetries. The results are presented in Fig.~\ref{fig:CK_symmetries}. In the left part of the diagram, quadrupoles are positionally arranged in pentagonal quasi-pinwheel structures around icosahedral vertices, with the spacing between the nearest neighbors increasing as we move towards the right. Fast decay of the quadrupolar interaction strength with the distance makes rotational directions for these pinwheels nearly independent of one-another which leads to configurations with lower symmetries (i.e. different directions of pinwheel rotations) having very similar energies to the icosahedral case. 

As we move towards the right and the quasi-pinwheels become closer to one another, their mutual interactions become significant and $C_2$, $C_3$ or $C_5$ configurations show lower energies compared to the icosahedral structures. Nevertheless, these still prevail along the top edge of the fundamental domain as we transition towards locally triangular arrangement of particles in the top right part of the diagram. This positional order also supports tetrahedral ground state configurations that occupy part of the top right corner of the fundamental domain. Conversely, ground state configurations in the center and lower right parts of the fundamental domain only show axial symmetries. In the bottom right corner, each quadrupole only has one nearest neighbor, leading to a strong preference for the local orthogonal T-configuration. This effectively precludes tetrahedral and icosahedral configuration symmetry as it is not compatible with $C_2$ rotations of the lattice. In the center of the diagram, many ground state configurations show no symmetries. This is attributed to strong geometric frustration as argued already for the case of the lattice C. 

In general, the symmetry diagram of quadrupolar ground states on CK lattices qualitatively strongly resembles the one for the dipolar case studied in Ref.~\cite{Copar2020}. Icosahedral and tetrahedral configurations are found in the same parts of the fundamental domain (even though they cover slightly larger regions in the quadrupolar case) and there are almost no configurations with dihedral symmetries in the ground state. Nevertheless, the locations of no symmetry regions are different between the dipolar and quadrupolar case, and in the bottom right corner, the $C_5$ and $C_3$ regions are exchanged which is a direct consequence of different preferential orientational configurations for pairs of dipoles and quadrupoles. 

\begin{figure}[!ht]
	\centering
	\includegraphics[width=0.6\linewidth]{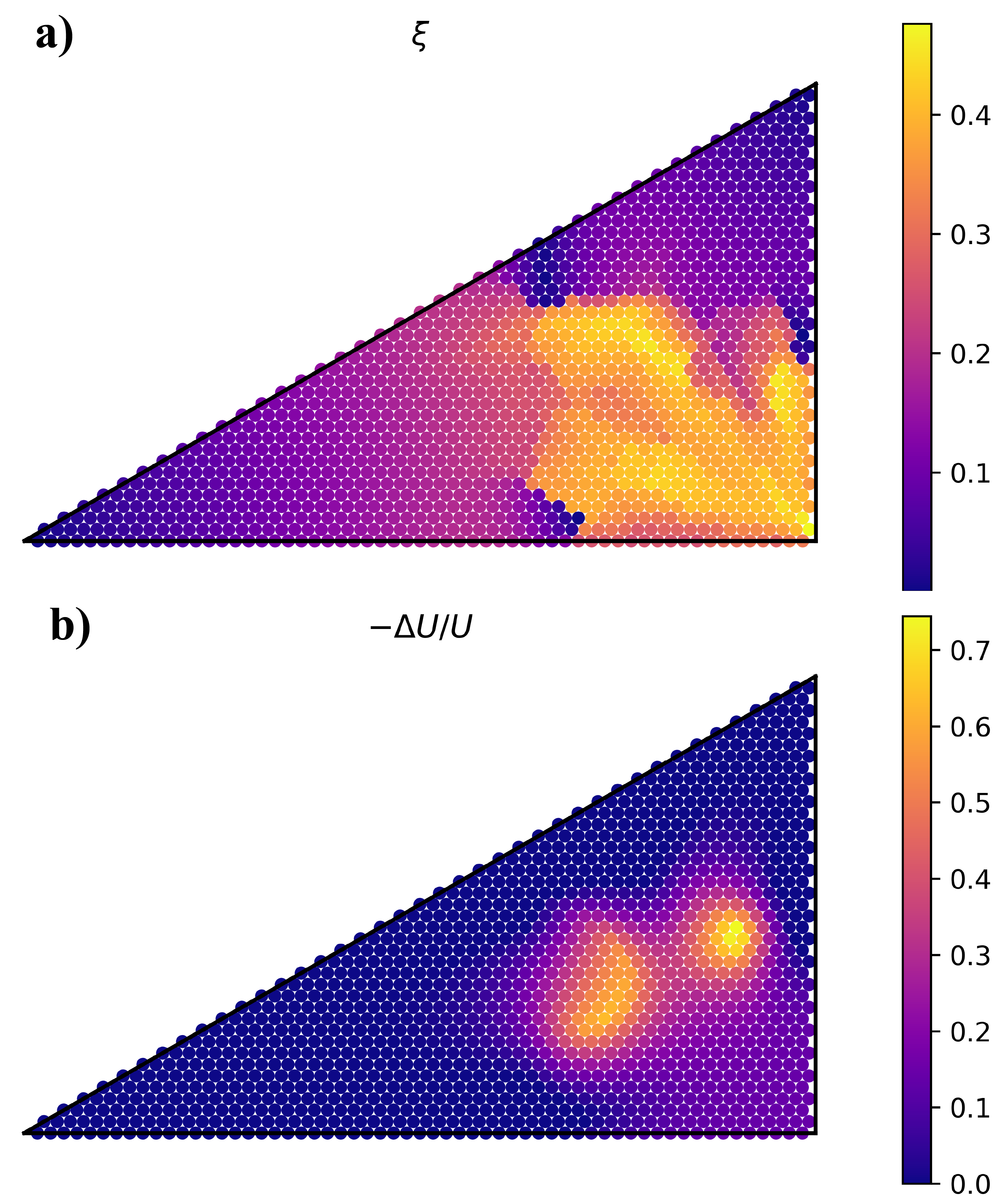}
	\caption{The out-of-plane parameter $\xi$ and the deviation $-\Delta \mathcal{U} / \mathcal{U}$ of the ground state energy from the icosahedral configuration energy, evaluated for the whole fundamental positional domain of CK lattices. Quadrupolar states on lattices in left and top right corners of the fundamental domain are approaching purely tangential ordering, with inclinations increasing towards the middle of the diagram. The configurations in bottom right corner and in the center exhibit higher values of $\xi$ due to local pair ordering and geometric frustration, respectively. Consequently, the energies of forced icosahedral orientational configurations are also higher in these regions, resulting in higher differences to the ground state energies.}
	\label{fig:CK_param}
\end{figure}

Returning back to the quadrupolar case, we plot the values of the out-of-plane parameter $\xi$ (Eq.~\ref{out-of-plane}) for ground state configurations in the whole fundamental domain (Fig.~\ref{fig:CK_param}a). The results in all three corners are as expected, approaching zero in the bottom left and top right corners, for configurations with pentagonal and triangular local arrangements of quadrupoles, respectively, and assuming higher values in the bottom right corner for configurations with pair T-ordering. Approaching the center of the fundamental domain from $\xi=0$ in the corners, we observe a gradual increase in $\xi$, with a more noticeable change towards higher values in the region with no ground state orientational order symmetry. Conversely, some configurations on the edges of the fundamental domain, where the positional order of the lattice is (nearly) equidistant (i.e. around lattices A and B as well as in the middle of the right edge), show close to tangential ordering.

Furthermore, we compare the energies of ground state configurations with the lowest energy icosahedral state on the same lattice. Note that the icosahedral configuration is not stable in general and may decay after the symmetry constraint is lifted, especially in the part of the diagram with non-symmetric ground states. The ratio $-\Delta \mathcal{U} / \mathcal{U}$ of the energy difference to the icosahedral state and the energy of the ground state configuration is shown in Fig.~\ref{fig:CK_param}b. It is small in the bottom left and top right parts of the fundamental domain which shows that reducing the symmetry to tetrahedral or axial does not lead to a substantial decrease in the configuration energy for locally pentagonal and triangular lattices. As expected, the energy difference is proportionally higher in the bottom right corner where quadrupoles order in local T-configurations, as well as in the center of the diagram with no symmetric ground states. There are two distinct peaks of the energy ratio; some configurations between the two also exhibiting symmetries in the ground state as shown in Fig.~\ref{fig:CK_symmetries}.

\subsection{Constrained solutions and different quadrupole tensors}\label{constr_sol}

In real systems, quadrupoles are usually not free to rotate but are rather constrained around some preferred tilt, dictated by the interaction with the substrate as well as temperature \cite{Ralser2016,Zottl2014}. Additionally, quadrupole eigentensors can differ from the rotationally symmetric linear type. We examine the dependence of the ground state configuration on the parameter $a$ that determines the quadrupolar eigenvalues, at different fixed inclinations $\theta$ between the surface normal and the eigenvector corresponding to the highest eigenvalue $e_1=2+a$. To do so, minimization over two Euler angles for each quadrupole is still required (only the third is constant and fixes the value of $\theta$).

\begin{figure}[!ht]
	\centering
	\includegraphics[width=\linewidth]{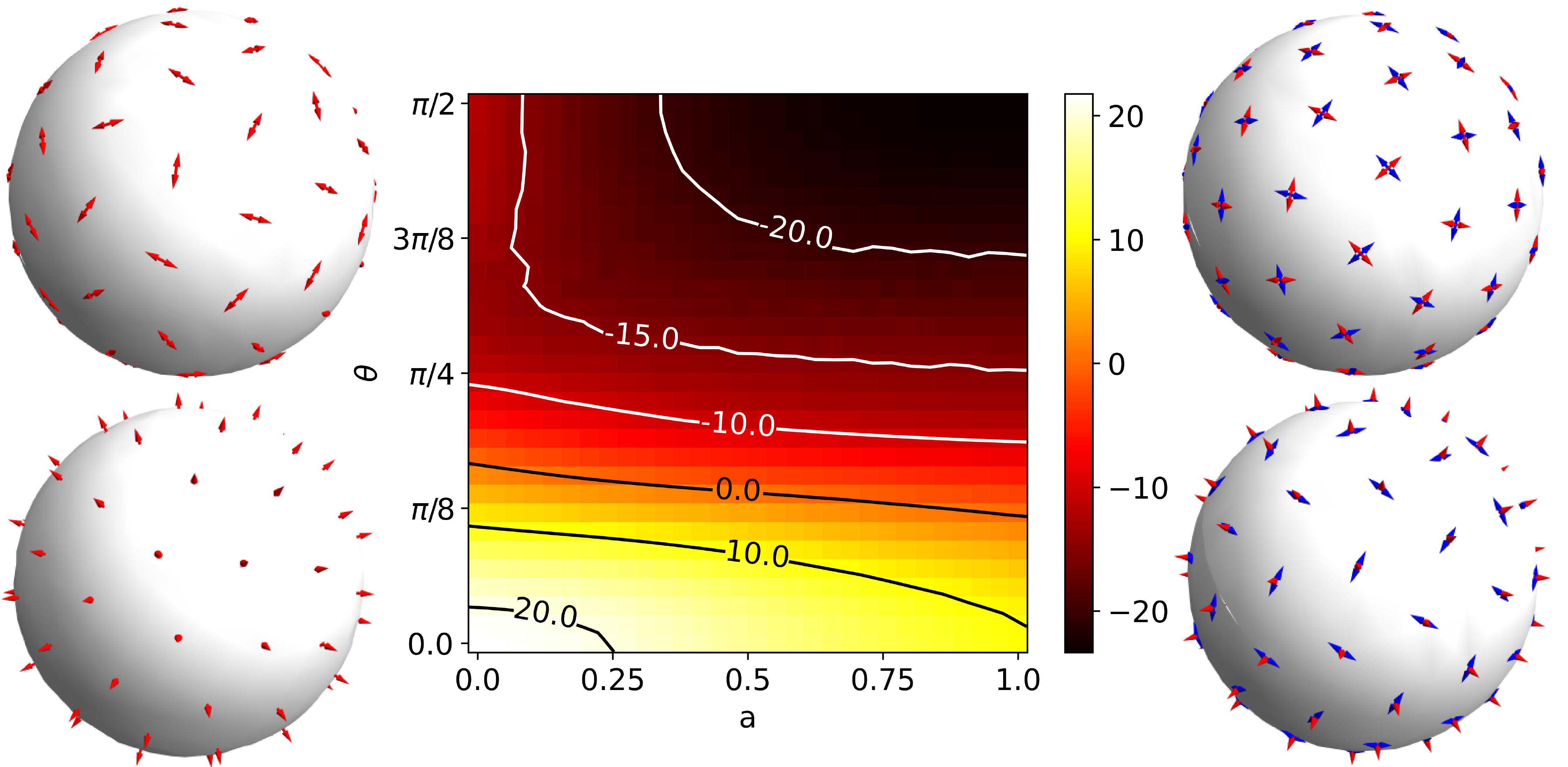}
	\caption{Energy heatmap (in units of $10^3$) along with a visual comparison of configurations between different quadrupole types for the $N=72$ Thomson lattice. Energies decrease with increasing the tilt constraint $\theta$ and the eigenvalue parameter $a$. The configuration at $(0, 0)$ is trivially radial and at $(1, 0)$ and $(0, \pi/2)$, we observe equivalent herringbone order for the main (red) and secondary (blue) axes, respectively. At $(1, \pi/2)$ herringbone order is also observed if we only consider any one of the two quadrupole axes.}
	\label{fig:quad_tan_heatmap}
\end{figure}

Figure~\ref{fig:quad_tan_heatmap} shows a heatmap of the optimal configuration energies in the $(a, \theta)$ plane for a quadrupolar system on the $N=72$ icosahedral Thomson lattice. Configurations at higher $\theta$ on the top of the diagram always show lower energies compared to lower $\theta$ configurations. Similarly, we observe a general decrease in energy as quadrupoles transition from the linear (left edge) to the planar (right edge) quadrupole type. Calculating energy heatmaps for quadrupolar systems on lattices of different sizes and types (including the CK lattices), we note that the dependence of energy on parameters $a$ and $\theta$ is qualitatively the same. We also visualize the limiting configurations in all four edges of the diagram. The configuration at $(0, 0)$ is trivially radial whereas at $(0, \pi/2)$, quadrupoles form regions with the herringbone orientational order as is also typical on 2D triangular lattices with forced tangential order. At the pentagonal lattice defects, the herringbones change orientational direction. Ordering at $(1, 0)$ is similar, as constraining one axis of planar quadrupoles to the radial direction results in the other axis lying tangent to the surface. The solution at $(1, \pi/2)$ is completely planar with neighboring quadrupoles trying to align perpendicular to one another. Considering only the main (red) axes, we again notice the local herringbone structures.

While Fig.~\ref{fig:quad_tan_heatmap} only shows ground state configurations for the limiting cases of the parametrization, the way in which configurations transition from perpendicular to tangential main axis orientation ($\theta=0\rightarrow\theta=\pi/2$) warrants an additional comment. If we consider only the orientations of the tangential components of main axis orientations at increasing inclinations ($\theta=0$ towards $\theta=\pi/2$), we notice that for small values of $\theta$, the main axes projections form a quasi-macrovortex. Around $\theta=\pi/4$, the projected orientations transition towards the herringbone order that then consistently emerges for $\theta \gtrsim \pi/3$. This behavior does not depend on the eigenvalues of the quadrupoles (parameter $a$). Additionally, such orientational order for all configurations of constrained quadrupoles also breaks any rotational symmetry that might emerge for a system of unconstrained linear quadrupoles on the same lattice.

\begin{figure}[!ht]
	\centering
	\includegraphics[width=0.7\linewidth]{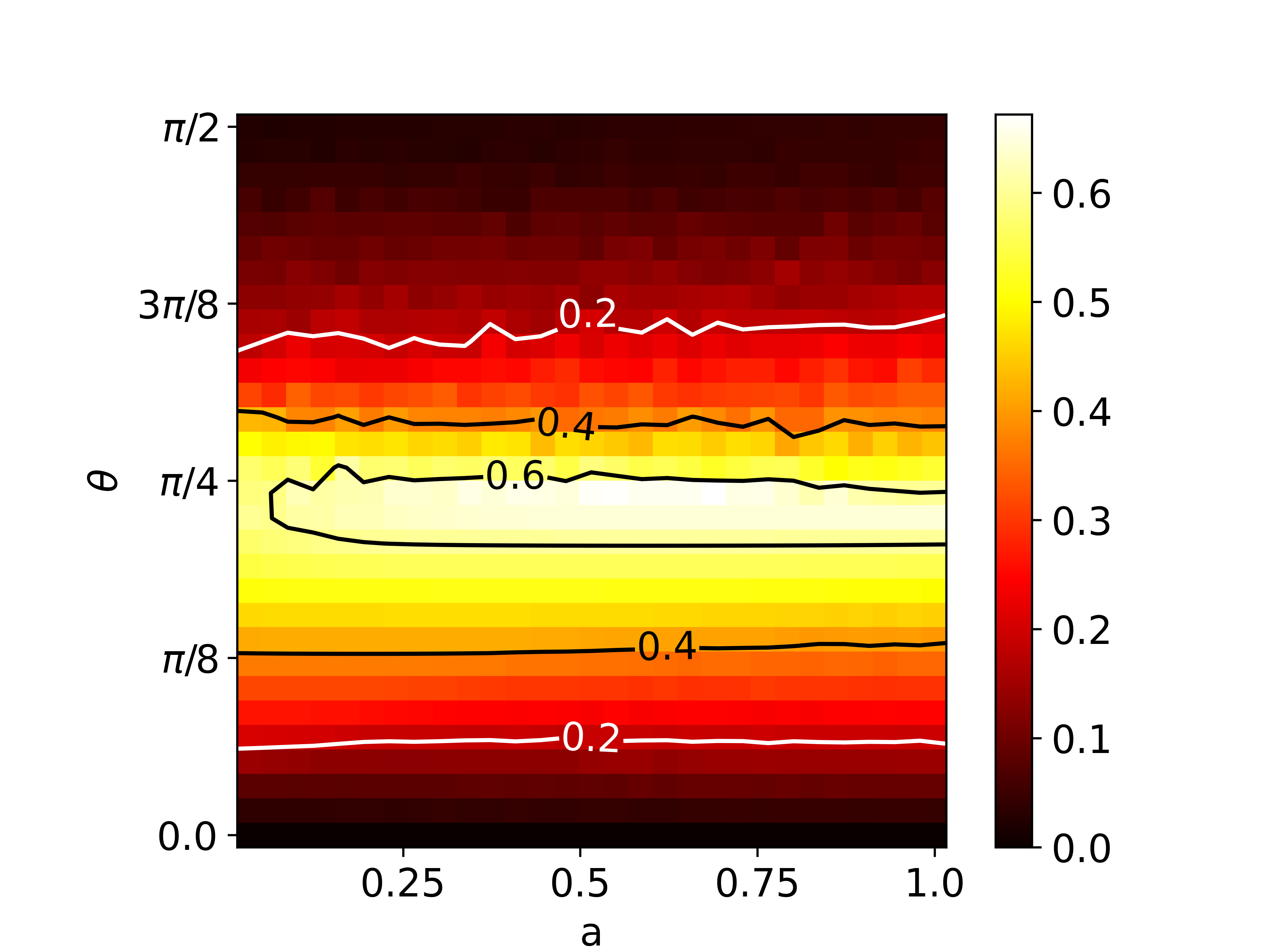}
	\caption{A heatmap of the average secondary axis inclination $\nu$ for the $N=72$ Thomson lattice on the whole parametric domain. The values at the bottom edge are zero by definition and the results for $a=0$ are omitted from the plot as they are ambiguous for linear quadrupoles. The inclinations show no dependence on $a$, they peak around $\theta=\pi/4$ with $\nu\approx 0.65$ and are close to zero for $\theta=\pi/2$, indicating that the fully planar quadrupolar configurations at this angle are energetically favorable.}
	\label{fig:quad_tan_inc}
\end{figure}

It should be noted that while we constrained the tilt of the main quadrupole axes (red double arrows in Fig.~\ref{fig:quad_tan_heatmap}) with respect to the surface normal, the orientations of the secondary axes (blue double arrows) are determined by energy minimization and can therefore be tilted out of the tangent plane. We plot the average inclination of the secondary axes $\hat{\bm{n}}_i$ to the surface normal $\nu = \sum_i |\hat{\bm{n}}_i \cdot \hat{\bm{r}}_i| / N$ in Fig.~\ref{fig:quad_tan_inc}. The result $\nu=0$ along the bottom edge of the diagram is trivial, as the secondary axis must be perpendicular to the main axis at $a>0$. Along the top edge, there is no constraint for the orientations of the secondary axes, therefore, values of $\nu$ close to 0 show that planar ordering is energetically favorable. Conversely, values of $\nu$ increase as we approach $\theta=\pi/4$ in the middle of the diagram. These configurations have both the main and secondary axis strongly tilted from the planar configuration with virtually no dependence on the eigenvalue parameter $a$ (for $a > 0$). The results for other lattices (including CK lattices) are similar.

\begin{figure}[!ht]
	\centering
	\includegraphics[width=\linewidth]{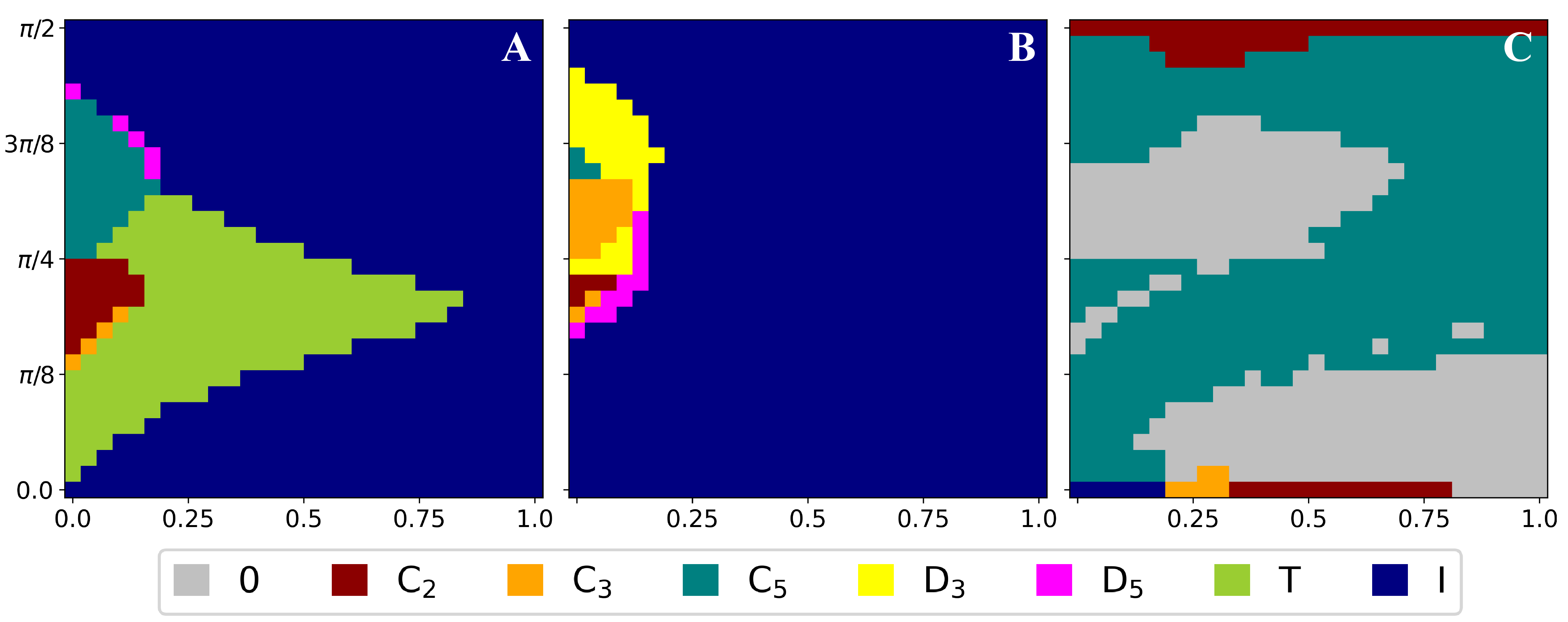}
	\caption{Ground state configuration symmetries for CK lattices A, B and C in dependence to the main axis inclination $\theta$ and eigenvalue parameter $a$ (axes are the same as in Fig.~\ref{fig:quad_tan_heatmap} and \ref{fig:quad_tan_inc}). On the lattices A and B, the most common symmetry is the icosahedral symmetry as in the case of freely rotating linear quadrupoles. The lattice A also shows a large tetrahedral region as well as smaller $C_3$ and $C_5$ regions. The $C_2$ and $D_5$ configurations appear to be only transitional. The results on the lattice B are even more diverse in the non-icosahedral part of the diagram, indicating possible minimization imperfections. The lattice C which does not support asymmetric ground state without the constraint to the main quadrupole axes inclinations, assumes the $C_5$ symmetry in a large portion of the parameter space.}
	\label{fig:CK_param_sym}
\end{figure}

We argue that locally triangular lattices do not support symmetric configurations of quadrupoles if we constrain their inclinations to $\theta>0$ or if we consider quadrupoles that deviate from the linear type ($a>0$). Geometric frustration that emerges on these lattices, especially as a consequence of lattice disclinations, cannot be offset by different quadrupole tilts as is the case for the symmetric ground states in Fig.~\ref{fig:quad_sym}, leading to a lack of long-range orientational order. In comparison, quadrupoles in many ground state configurations on CK lattices already have the same inclinations and, due to vacancies in the positional order, local herringbone structures do not form on these lattices. This opens up the possibility that long-range ordered symmetric quadrupolar configurations exist on CK lattices also if the orientations are constrained to a specific inclination $\theta$. More specifically, this inclination, as well as the type of the quadrupole given by parameter $a$, can influence the symmetry of the configuration as shown in Fig.~\ref{fig:CK_param_sym} for the equidistant CK lattices A, B and C.

Systems on lattices A and B show icosahedral ground state structures in the majority of the diagram. In both cases, this symmetry breaks along the left edge, i.e. as we increase the tilts of the quadrupoles with respect to the surface normal. For the lattice A, the ground state first transforms into the tetrahedral configuration that also prevails towards the center of the diagram (with increasing $a$) and, increasing $\theta$ even further, we encounter regions with $C_3$ and $C_5$ symmetries. Notice also the narrow $C_2$ and $D_5$ transitional regions. For the lattice B, the part of the diagram without the icosahedral symmetry is subdivided into many smaller regions, possibly indicating that we failed to find the true ground state in all points. Similarly, some imperfections in the simulations are present in the diagram of the lattice C. Despite the lack of symmetry in the ground state of freely rotating linear quadrupoles (Fig.~\ref{fig:CK_symmetries}) on this lattice, a large part of the diagram for the constrained case shows $C_5$ symmetric ordering. Both inclination extremes, ie. the radial ($\theta=0$) and tangential cases ($\theta=\pi/2$), assume configurations with different symmetries.

\section{Conclusions}

We investigated the role of local positional order, lattice symmetry, and inclination constraints on the orientational ordering in quadrupolar systems on the surface of the sphere. Triangular local positional order in particular leads to geometric frustration and in general does not support global (long-range) orientational order, as indicated by the solutions on many Thomson and Tammes lattices. However, we have shown that lattice symmetry can stabilize long range order for certain lattices, resulting in quadrupolar configurations with different axial symmetries as well as pure pinwheel configurations with higher symmetry, i.e. tetrahedral, octahedral and icosahedral. Orientational order supported by positional symmetry emerges also in many ground state configurations on CK lattices. The symmetry diagram shows rich behavior that can be directly attributed to the competition between the local geometric frustration and icosahedral lattice symmetry. We further expended the discussion by examining ordering of general quadrupole tensors. Additionally, we constrained the inclinations of the main quadrupole axis to model possible effects of interactions with the substrate. We find that both the inclination and the type of quadrupoles influence the ground state ordering, in particular its symmetry. 

In real spherical quadrupolar systems, such as the adsorbates of nonpolar molecules on spherical fullerenes, orientational ordering is usually influenced by other interactions besides the quadrupole-quadrupole interaction. Most notably, interactions with the substrate can impose preferential tilt of the quadrupoles, leading to some discrepancies with our results. Nevertheless, by keeping our discussion as general as possible, the role of quadrupolar interaction on orientational ordering can be analyzed and evaluated in a diverse range of physical systems. Further investigations could focus on a generalization to more complex interparticle and particle-substrate interactions, or modeling thermal behaviour of the system at a finite temperature, in turn improving the agreement of the simulation results with the behavior in real systems. Additionally, our results show the possibility for the construction of systems with the desired orientational order by choosing the appropriate positional lattice and manipulating the inclinations of the quadrupoles. Finally, in physical systems that do not fix the positional order of the quadrupolar units, joint minimization of positional and orientational interaction free energy would be required, subverting the role of the pre-imposed lattice symmetries. We leave this discussion open for future investigation.

\section*{Acknowledgments}
We thank A. Božič for helpful discussions and suggestions during our research.
We acknowledge support by Slovenian Research Agency (ARRS) under contracts P1-0099 and J1-9149. The work is associated with the COST action CA17139.

\bibliographystyle{science}
\bibliography{bibliography}

\begin{thebibliography}{10}

\bibitem{Vedmedenko2008}
E.~Y. Vedmedenko, N.~Mikuszeit, {\it {ChemPhysChem}\/} {\bf 9}, 1222 (2008).

\bibitem{Shen2003}
Y.~R. Shen, {\it Principles of nonlinear optics\/} (Wiley, 2003).

\bibitem{Buckingham1967}
A.~D. Buckingham, {\it Adv.~Chem.~Phys.\/} (John Wiley {\&} Sons, Inc., 1967),
  pp. 107--142.

\bibitem{Birnbaum1985}
G.~Birnbaum, {\it Phenomena induced by intermolecular interactions\/} (Plenum
  Press, 1985).

\bibitem{Vedmedenko2005}
E.~Y. Vedmedenko, N.~Mikuszeit, H.~P. Oepen, R.~Wiesendanger, {\it Phys. Rev.
  Lett.\/} {\bf 95}, 207202 (2005).

\bibitem{Ewerlin2013}
M.~Ewerlin, {\it et~al.\/}, {\it Phys. Rev. Lett.\/} {\bf 110}, 177209 (2013).

\bibitem{Mikuszeit2009}
N.~Mikuszeit, {\it et~al.\/}, {\it Phys. Rev. B\/} {\bf 80}, 014402 (2009).

\bibitem{Berlinsky1978}
A.~J. Berlinsky, A.~B. Harris, {\it Phys.~Rev.~Lett.\/} {\bf 40}, 1579 (1978).

\bibitem{OShea1979}
S.~F. O'Shea, M.~L. Klein, {\it Chem.~Phys.~Lett.\/} {\bf 66}, 381 (1979).

\bibitem{OShea1982}
S.~F. O'Shea, M.~L. Klein, {\it Phys.~Rev.~ B\/} {\bf 25}, 5882 (1982).

\bibitem{Mouritsen1982}
O.~G. Mouritsen, A.~J. Berlinsky, {\it Phys.~Rev.~ Lett.\/} {\bf 48}, 181
  (1982).

\bibitem{Pan1982}
R.~P. Pan, R.~D. Etters, K.~Kobashi, V.~Chandrasekharan, {\it J.~Chem.~Phys.\/}
  {\bf 77}, 1035 (1982).

\bibitem{Peters1985}
C.~Peters, M.~L. Klein, {\it Mol.~Phys.\/} {\bf 54}, 895 (1985).

\bibitem{Kim1999}
K.~Kim, N.~S. Sullivan, {\it J.~Low~Temp.~Phys.\/} {\bf 114}, 173 (1999).

\bibitem{Sullivan2000}
N.~S. Sullivan, K.~Kim, {\it J.~Low~Temp.~Phys.\/} {\bf 120}, 89 (2000).

\bibitem{Boyd2002}
D.~Boyd, F.~Hess, G.~Hess, {\it Surf.~Sci.\/} {\bf 519}, 125 (2002).

\bibitem{Ustinov2016}
E.~A. Ustinov, {\it Carbon\/} {\bf 100}, 52 (2016).

\bibitem{Ustinov2018}
E.~Ustinov, V.~Gorbunov, S.~Akimenko, {\it J.~Phys.~Chem.~C\/} {\bf 122}, 2897
  (2018).

\bibitem{Hamida1994}
J.~A. Hamida, N.~S. Sullivan, M.~D. Evans, {\it Phys.~Rev.~Lett.\/} {\bf 73},
  2720 (1994).

\bibitem{Raugei1997}
S.~Raugei, G.~Cardini, V.~Schettino, H.~Jodl, {\it Chem.~Phys.\/} {\bf 106},
  8196 (1997).

\bibitem{Sullivan1978}
N.~S. Sullivan, M.~Devoret, B.~P. Cowan, C.~Urbina, {\it Phys.~Rev.~B\/} {\bf
  17}, 5016 (1978).

\bibitem{Sullivan1988}
N.~S. Sullivan, {\it Can.~J.~Chem.\/} {\bf 66}, 908 (1988).

\bibitem{Kim1997}
K.~Kim, N.~S. Sullivan, {\it Phys.~Rev.~B\/} {\bf 55}, R664 (1997).

\bibitem{Harris1984}
A.~B. Harris, O.~G. Mouritsen, A.~J. Berlinsky, {\it Can.~J.~Phys.\/} {\bf 62},
  915 (1984).

\bibitem{Klenin1982}
M.~A. Klenin, S.~F. Pate, {\it Phys.~Rev.~B\/} {\bf 26}, 3969 (1982).

\bibitem{Mederos1990}
L.~Mederos, E.~Chac{\'{o}}n, P.~Tarazona, {\it Phys.~Rev.~ B\/} {\bf 42}, 8571
  (1990).

\bibitem{Massidda1984}
V.~Massidda, J.~Hernando, {\it Physica A\/} {\bf 128}, 318 (1984).

\bibitem{Marx1993}
D.~Marx, O.~Opitz, P.~Nielaba, K.~Binder, {\it Phys.~Rev.~Lett.\/} {\bf 70},
  2908 (1993).

\bibitem{Bowick2009}
M.~J. Bowick, L.~Giomi, {\it Adv. Phys.\/} {\bf 58}, 449 (2009).

\bibitem{Vitelli2006}
V.~Vitelli, D.~R. Nelson, {\it Phys. Rev. E\/} {\bf 74}, 021711 (2006).

\bibitem{Li2013}
Y.~Li, H.~Miao, H.~Ma, J.~Z.~Y. Chen, {\it Soft Matter\/} {\bf 9}, 11461
  (2013).

\bibitem{Sknepnek}
R.~Sknepnek, S.~Henkes, {\it Phys. Rev. E\/} {\bf 91}, 022306 (2015).

\bibitem{Praetorius}
S.~Praetorius, A.~Voigt, R.~Wittkowski, H.~Löwen, {\it Phys. Rev. E\/} {\bf
  97}, 70 (2018).

\bibitem{Kravchuk2016}
V.~P. Kravchuk, {\it et~al.\/}, {\it Phys. Rev. B\/} {\bf 94}, 144402 (2016).

\bibitem{Milagre2007}
G.~Milagre, W.~A. Moura-Melo, {\it Phys.~Lett.~A\/} {\bf 368}, 155 (2007).

\bibitem{Sloika2017}
M.~I. Sloika, D.~D. Sheka, V.~P. Kravchuk, O.~V. Pylypovskyi, Y.~Gaididei, {\it
  J.~Magn.~Magn.~Mater\/} {\bf 443}, 404 (2017).

\bibitem{Gnidovec2020}
A.~Gnidovec, S.~Čopar, {\it Phys. Rev. B\/} {\bf 102}, 075416 (2020).

\bibitem{Copar2020}
S.~Čopar, A.~Božič, {\it Phys. Rev. Research\/} {\bf 2}, 043199 (2020).

\bibitem{Vedmedenko2008_1}
E.~Vedmedenko, S.~E.-D. Mandel, R.~Lifshitz, {\it Philos.~Mag.\/} {\bf 88},
  2197 (2008).

\bibitem{MartnezAlonso2000}
A.~Mart{\'{\i}}nez-Alonso, J.~M.~D. Tasc{\'{o}}n, E.~J. Bottani, {\it
  Langmuir\/} {\bf 16}, 1343 (2000).

\bibitem{Ralser2016}
S.~Ralser, {\it et~al.\/}, {\it Phys.~Chem.\/} {\bf 18}, 3048 (2016).

\bibitem{Zottl2014}
S.~Z\"{o}ttl, {\it et~al.\/}, {\it Carbon\/} {\bf 69}, 206 (2014).

\bibitem{Vedmedenko2007}
E.~Y. Vedmedenko, {\it Competing interactions and patterns in nanoworld\/}
  (Wiley-VCH, 2007).

\bibitem{Thorne1980}
K.~S. Thorne, {\it Rev. Mod. Phys.\/} {\bf 52}, 299 (1980).

\bibitem{Franzini2018}
S.~Franzini, L.~Reatto, D.~Pini, {\it Soft Matter\/} {\bf 14}, 8724 (2018).

\bibitem{Bozic2019}
A.~Lošdorfer~Božič, S.~Čopar, {\it Phys. Rev. E\/} {\bf 99}, 032601 (2019).

\bibitem{SuppMat}
See Supplemental Material at \textit{url} for more details on quadrupolar
  configurations at different system sizes, including on the Tammes and
  Fibonacci lattices.

\bibitem{Calvo2012}
F.~Calvo, {\it Phys.~Rev.~B\/} {\bf 85}, 060502 (2012).

\bibitem{Echt2013}
O.~Echt, {\it et~al.\/}, {\it {ChemPlusChem}\/} {\bf 78}, 910 (2013).

\bibitem{Kaiser2013}
A.~Kaiser, {\it et~al.\/}, {\it J.~Chem.~Phys.\/} {\bf 138}, 074311 (2013).

\bibitem{Bruinsma2015}
R.~F. Bruinsma, W.~S. Klug, {\it Annu. Rev. Condens. Matter Phys.\/} {\bf 6},
  245 (2015).

\bibitem{Rochal2017}
S.~B. Rochal, O.~V. Konevtsova, V.~L. Lorman, {\it Nanoscale\/} {\bf 9}, 12449
  (2017).

\end{thebibliography}

\end{document}

\title{\textbf{Long-range order in quadrupolar systems on spherical surfaces -- SUPPLEMENTAL MATERIAL}}
\author{Andraž Gnidovec and Simon Čopar \\ \textit{University of Ljubljana, Faculty of Mathematics and Physics, SI-1000 Ljubljana}, Slovenia}
\date{}

\maketitle

\section{Additional results}

In the main article, we showed several examples of ordering in quadrupolar systems on different Thomson lattices. We argue that the local arrangement of quadrupoles as well as global positional symmetry both contribute to the resulting quadrupolar order. As the Thomson lattice represents only one way of uniformly distributing points on a sphere, we further look to confirm this by exploring the quadrupolar states on the Tammes and Fibonacci lattices. The Tammes lattice is locally triangular and thus structurally similar to the Thomson lattice, whereas the Fibonacci lattice\footnotemark becomes locally square around the equator as we increase the system size.

\footnotetext{We parametrize the Fibonacci lattice of size $N$ in spherical coordinates as
	\begin{equation*}
	(\theta_k, \phi_k) = \left(\arccos \left[\frac{2k}{N + 1} - 1\right], 2\pi(2 - \alpha)k \right),
	\end{equation*}
	where $\alpha = \tfrac{\sqrt{5}-1}{2}$ is the golden ratio and $k=1\ldots N$. Different definitions are possible but they result in a structurally similar lattice.}

\begin{figure*}[!hb]
	\centering
	\includegraphics[width=0.9\linewidth]{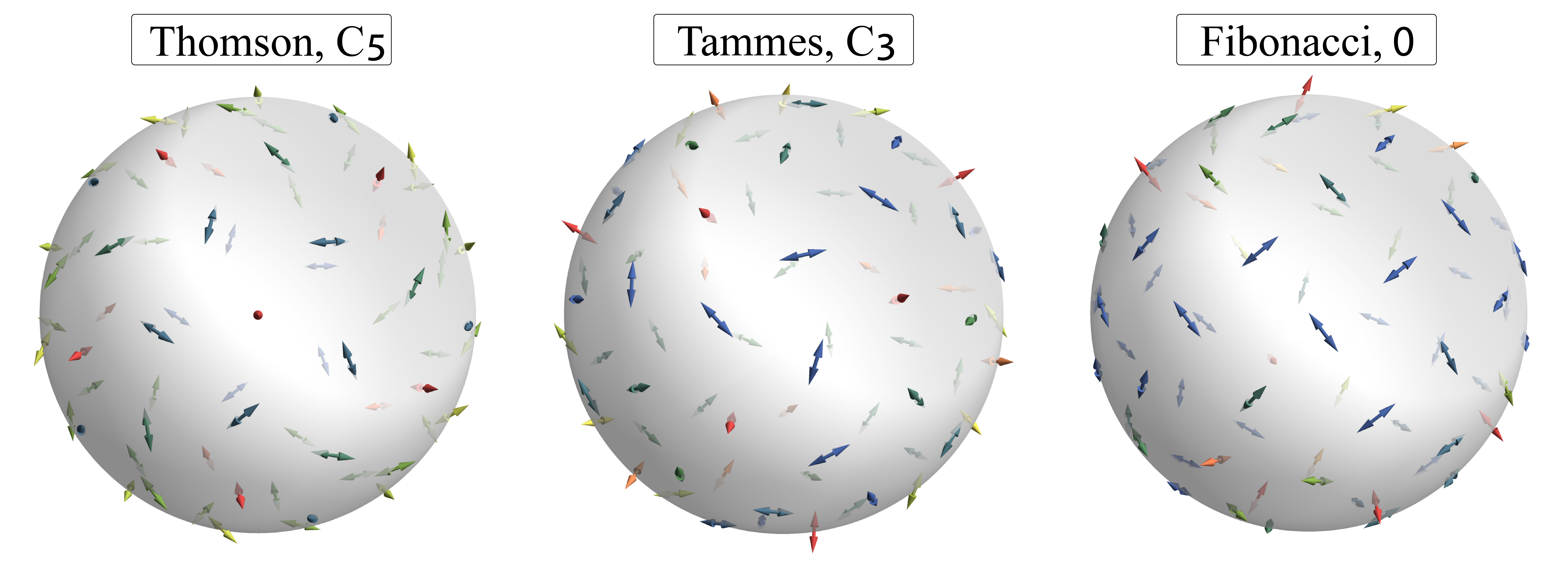}
	\caption{Comparison of configurations on different spherical lattices at system size $N=72$.}
	\label{fig:comparison}
\end{figure*}

A comparison of configurations at $N=72$ for different lattices is presented in Fig.~\ref{fig:comparison}. The ground state on the icosahedral Thomson lattice shows $C_5$ symmetry. Conversely, the Tammes lattice, while similar, lacks the high positional symmetry of the Thomson case, resulting in a ground state without any symmetries. The configuration shown in Fig.\ref{fig:comparison} is the only symmetric excited state and belongs to $C_3$ point group. 
Comparing Thomson and Tammes configurations, we observe in both cases the tendency of quadrupoles to form pinwheel structures, at least in the vicinity of the symmetry axis. The situation differs substantially when we look at the configuration on the Fibonacci lattice. Locally square distribution of points around the equator results in the formation of windmill structures, similar to what we observe on planar square lattice. Around the poles, the lattice is highly irregular and long-range order is absent.

\begin{figure*}[!ht]
	\centering
	\includegraphics[width=0.7\linewidth]{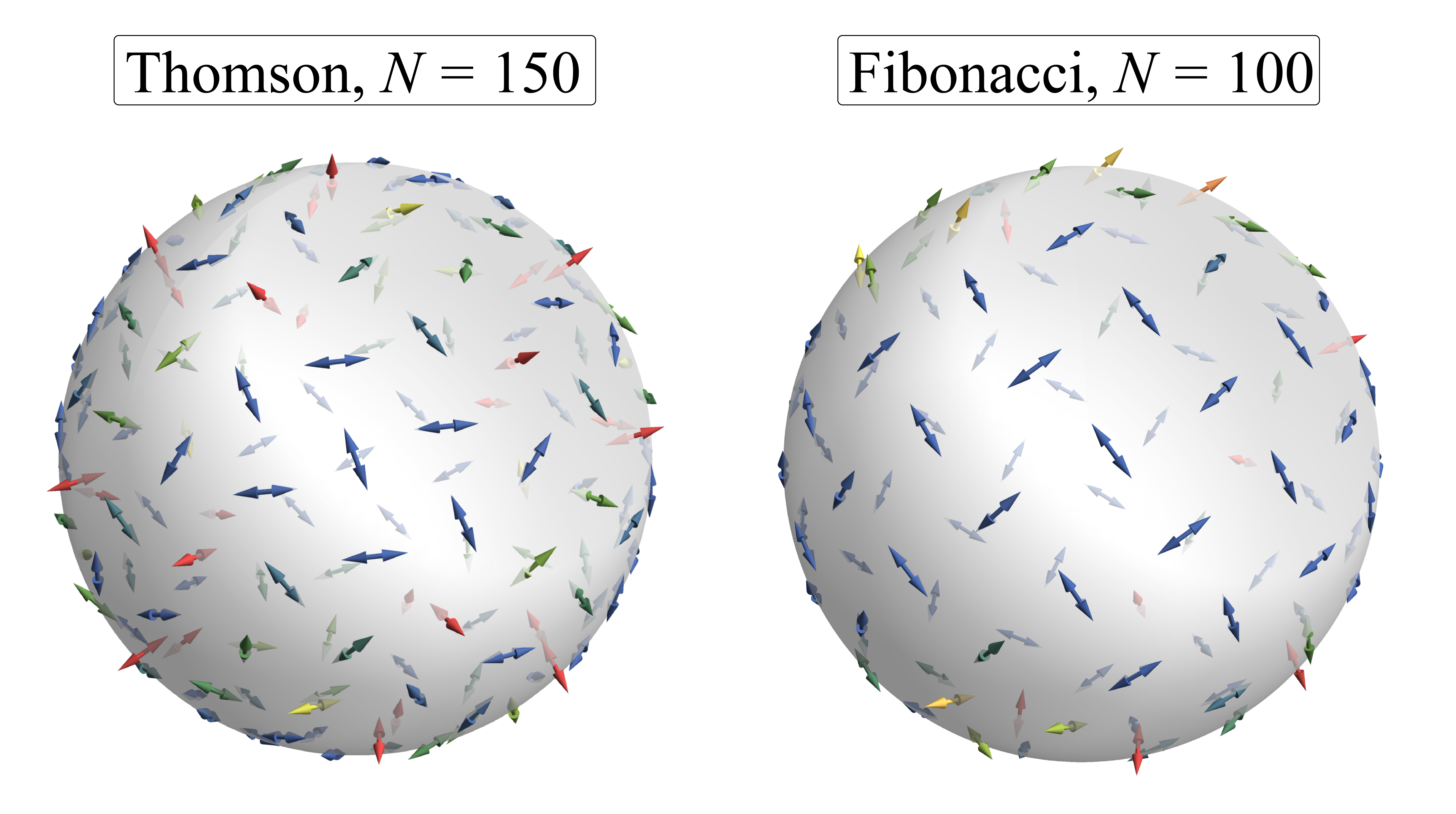}
	\caption{Examples of Thomson and Fibonacci quadrupolar ground states at higher system sizes.}
	\label{fig:highN}
\end{figure*}

The choice of using a deterministic minimization method for investigating quadrupolar systems on a sphere limits the maximal system size we are able to explore to $N\approx150$. Nevertheless, we argue that this is enough to discern ordering trends in bigger systems. Figure~\ref{fig:highN} shows ground state configurations for $N=150$ tetrahedral Thomson lattice and $N=100$ Fibonacci lattice. The former shows emergent pinwheel structures over the whole surface of the sphere, however, the arrangement of lattice defects does not support higher symmetry ordering (as e.g. in the case for $N=122$) and the configuration only shows $C_2$ symmetry. Nevertheless, the configuration supports our assumption that in the limit of high $N$, quadrupolar systems on locally triangular lattices transition towards a pinwheel ground state (away from lattice defects) as the role of curvature decreases. This is further corroborated by lower values of out-of-plane parameter $\xi$ for all configurations found at this system size ($0.30 \lesssim \xi \lesssim 0.38$, ground state value $\xi_{GS}=0.303$) compared to values of $\xi$ at smaller $N$. The Fibonacci configuration at $N=100$ shows similar behavior to the $N=72$ case in Fig.~\ref{fig:comparison} but with a much wider windmill region. This results in the out-of-plane parameter value of $\xi=0.198$ which is smaller than the pinwheel-limited value of 1/4 for locally triangular lattices (in the limit of high system sizes).

\section{Tables of optima}

Below, we list lowest energy configuration results, along with configurations with the highest symmetry at given $N$, for systems of linear quadrupoles both the Thomson and Tammes lattices. The energy values correspond to the quadrupole tensor eigenvalues that satisfy $\text{Tr}(Q^2) = 1$.

\begin{table*}[!ht]
	\vspace{-3ex}
	\begin{subtable}[t]{0.5\linewidth}
		\begin{tabular}[t]{c|cccc} 
			$N$ & $E_{GS}$ & PG$_{GS}$ & $E_{sym}$ & PG$_{sym}$ \\ \hline\hline 
			10 & -17.94614 & $0$ & -14.79685 & $C_2$\\ \hline 
			11 & -25.87675 & $0$ & -24.98204 & $C_2$\\ \hline 
			12 & -30.76187 & $S_6$ & -30.76187 & $S_6$\\ \hline 
			13 & -44.19957 & $0$ & -44.19957 & $0$\\ \hline 
			14 & -55.17387 & $0$ & -47.42441 & $C_6$\\ \hline 
			15 & -71.38314 & $0$ & -69.48079 & $C_3$\\ \hline 
			16 & -85.16844 & $0$ & -81.26653 & $T$\\ \hline 
			17 & -115.1684 & $C_2$ & -115.1684 & $C_2$\\ \hline 
			18 & -132.0108 & $C_2$ & -132.0108 & $C_2$\\ \hline 
			19 & -165.4244 & $0$ & -164.6409 & $C_2$\\ \hline 
			20 & -190.3918 & $0$ & -181.4474 & $C_2$\\ \hline 
			21 & -231.9683 & $C_2$ & -231.9683 & $C_2$\\ \hline 
			22 & -266.2751 & $S_4$ & -266.2751 & $S_4$\\ \hline 
			23 & -311.1626 & $0$ & -304.6335 & $C_2$\\ \hline 
			24 & -350.321 & $C_3$ & -350.0883 & $C_4$\\ \hline 
			25 & -425.6512 & $0$ & -425.6512 & $0$\\ \hline 
			26 & -475.3702 & $0$ & -447.0872 & $C_2$\\ \hline 
			27 & -553.6096 & $C_5$ & -553.6096 & $C_5$\\ \hline 
			28 & -595.3587 & $0$ & -580.0058 & $C_2$\\ \hline 
			29 & -696.4318 & $0$ & -689.0495 & $C_2$\\ \hline 
			30 & -780.5806 & $0$ & -772.0964 & $C_2$\\ \hline 
			31 & -852.7957 & $0$ & -842.0158 & $C_3$\\ \hline 
			32 & -944.3894 & $S_{10}$ & -944.3894 & $S_{10}$\\ \hline 
			33 & -1095.938 & $0$ & -1095.938 & $0$\\ \hline 
			34 & -1227.507 & $C_2$ & -1205.856 & $D_2$\\ \hline 
			35 & -1387.788 & $0$ & -1387.788 & $0$\\ \hline 
			36 & -1507.733 & $0$ & -1427.541 & $C_2$\\ \hline 
			37 & -1629.9 & $C_5$ & -1629.9 & $C_5$\\ \hline 
			38 & -1899.729 & $S_{12}$ & -1899.729 & $S_{12}$\\ \hline 
			39 & -1923.571 & $C_2$ & -1896.799 & $C_3$\\ \hline 
			40 & -2077.554 & $0$ & -2024.524 & $C_3$\\ \hline 
			41 & -2266.805 & $C_2$ & -2266.805 & $C_2$\\ \hline 
			42 & -2483.26 & $0$ & -2420.833 & $C_2$\\ \hline 
			43 & -2685.682 & $0$ & -2685.682 & $0$\\ \hline 
			44 & -3153.106 & $O$ & -3153.106 & $O$\\ \hline 
			45 & -3203.02 & $0$ & -3192.192 & $C_2$\\ \hline 
			46 & -3433.877 & $C_2$ & -3296.245 & $D_2$\\ \hline 
			47 & -3780.18 & $0$ & -3780.18 & $0$\\ \hline 
			48 & -4099.089 & $C_4$ & -4056.577 & $D_4$\\ \hline 
			49 & -4377.138 & $0$ & -4232.584 & $C_3$\\ \hline 
			50 & -4722.145 & $S_{10}$ & -4722.145 & $S_{10}$\\ \hline 
			51 & -5071.913 & $0$ & -5071.913 & $0$\\ \hline 
			52 & -5311.184 & $C_3$ & -5311.184 & $C_3$\\ \hline 
			53 & -5771.033 & $0$ & -5641.892 & $C_2$\\ \hline 
			54 & -6121.877 & $0$ & -6001.246 & $C_2$\\ \hline 
			55 & -6527.448 & $0$ & -6490.042 & $C_2$\\ \hline 
			56 & -6869.911 & $0$ & -6740.875 & $C_2$\\ \hline 
			57 & -7390.209 & $C_3$ & -7390.209 & $C_3$\\ \hline 
		\end{tabular}
	\end{subtable}
	\begin{subtable}[t]{0.5\linewidth}
		\begin{tabular}[t]{c|cccc} 
			$N$ & $E_{GS}$ & PG$_{GS}$ & $E_{sym}$ & PG$_{sym}$ \\ \hline\hline 
			58 & -7847.213 & $C_2$ & -7847.213 & $C_2$\\ \hline 
			59 & -8421.899 & $0$ & -8292.696 & $C_2$\\ \hline 
			60 & -8729.22 & $0$ & -8707.275 & $C_3$\\ \hline 
			61 & -9427.155 & $0$ & -9427.155 & $0$\\ \hline 
			62 & -9841.105 & $0$ & -9675.032 & $C_5$\\ \hline 
			63 & -10541.75 & $0$ & -10497.81 & $C_2$\\ \hline 
			64 & -11218.07 & $0$ & -10629.97 & $C_2$\\ \hline 
			65 & -11800.14 & $0$ & -11696.37 & $C_2$\\ \hline 
			66 & -12593.85 & $0$ & -12593.85 & $0$\\ \hline 
			67 & -12792.58 & $0$ & -12666.86 & $C_2$\\ \hline 
			68 & -13687.36 & $0$ & -13250.64 & $C_2$\\ \hline 
			69 & -14379.16 & $C_3$ & -14379.16 & $C_3$\\ \hline 
			70 & -15305.46 & $S_4$ & -15305.46 & $S_4$\\ \hline 
			71 & -15937.29 & $0$ & -15937.29 & $0$\\ \hline 
			72 & -16265.76 & $C_5$ & -16265.76 & $C_5$\\ \hline 
			73 & -17581.62 & $0$ & -17217.24 & $C_2$\\ \hline 
			74 & -18796.39 & $0$ & -18796.39 & $0$\\ \hline 
			75 & -19333.59 & $0$ & -19188.43 & $C_3$\\ \hline 
			76 & -20496.73 & $0$ & -20496.73 & $0$\\ \hline 
			77 & -20785.71 & $C_5$ & -20785.71 & $C_5$\\ \hline 
			78 & -21890.22 & $C_2$ & -21675.97 & $S_6$\\ \hline 
			79 & -22887.72 & $0$ & -22887.72 & $0$\\ \hline 
			80 & -24140.24 & $0$ & -23691.7 & $C_4$\\ \hline 
			81 & -25375.87 & $0$ & -25166.6 & $C_2$\\ \hline 
			82 & -26399.32 & $0$ & -26314.82 & $C_2$\\ \hline 
			83 & -27664.05 & $0$ & -27376.7 & $C_2$\\ \hline 
			84 & -29065.02 & $0$ & -28927.9 & $C_2$\\ \hline 
			85 & -29901.36 & $0$ & -29554.66 & $C_2$\\ \hline 
			86 & -31580.65 & $0$ & -31580.65 & $0$\\ \hline 
			87 & -32441.53 & $0$ & -32272.87 & $C_2$\\ \hline 
			88 & -33681.07 & $0$ & -33681.07 & $0$\\ \hline 
			89 & -35199.35 & $0$ & -35006.52 & $C_2$\\ \hline 
			90 & -36214.78 & $0$ & -36214.78 & $0$\\ \hline 
			91 & -37825.44 & $0$ & -37636.89 & $C_2$\\ \hline 
			92 & -39411.5 & $0$ & -39411.5 & $0$\\ \hline 
			93 & -41005.54 & $0$ & -41005.54 & $0$\\ \hline 
			94 & -42724.3 & $C_2$ & -42724.3 & $C_2$\\ \hline 
			95 & -44286.94 & $0$ & -43545.95 & $C_2$\\ \hline 
			96 & -46302.03 & $0$ & -46302.03 & $0$\\ \hline 
			97 & -47523.83 & $0$ & -47523.83 & $0$\\ \hline 
			98 & -49221.05 & $0$ & -49221.05 & $0$\\ \hline 
			99 & -51178.49 & $0$ & -50711.02 & $C_2$\\ \hline 
			100 & -52712.89 & $0$ & -52638.11 & $C_3$\\ \hline 
			122 & -105055.6 & $I$ & -105055.6 & $I$\\ \hline 
			132 & -139470.9 & $C_5$ & -139470.9 & $C_5$\\ \hline 
			136 & -155127.9 & $0$ & -155127.9 & $0$\\ \hline 
			150 & -228044.5 & $C_2$ & -220168.9 & $D_2$\\ \hline 
		\end{tabular}
	\end{subtable}
	\caption{Ground state and highest symmetry solutions on the Thomson lattices.}
	\label{tab:Thomson}
\end{table*}

\begin{table*}[!ht]
	\begin{subtable}[t]{0.5\linewidth}
		\begin{tabular}[t]{c|cccc} 
			$N$ & $E_{GS}$ & PG$_{GS}$ & $E_{sym}$ & PG$_{sym}$ \\ \hline\hline 
			10 & -17.85698 & $0$ & -17.25249 & $C_2$\\ \hline 
			11 & -26.26433 & $0$ & -26.26433 & $0$\\ \hline 
			12 & -30.76187 & $S_6$ & -30.76187 & $S_6$\\ \hline 
			13 & -45.38164 & $0$ & -44.61766 & $C_4$\\ \hline 
			14 & -59.1275 & $C_2$ & -58.55624 & $S_4$\\ \hline 
			15 & -74.68377 & $0$ & -74.68377 & $0$\\ \hline 
			16 & -96.20829 & $S_8$ & -96.20829 & $S_8$\\ \hline 
			17 & -114.0142 & $C_2$ & -114.0142 & $C_2$\\ \hline 
			18 & -131.9526 & $0$ & -131.9526 & $0$\\ \hline 
			19 & -168.5474 & $0$ & -168.5474 & $0$\\ \hline 
			20 & -188.9719 & $0$ & -180.5959 & $C_2$\\ \hline 
			21 & -229.2873 & $0$ & -229.2873 & $0$\\ \hline 
			22 & -275.1527 & $0$ & -275.1527 & $0$\\ \hline 
			23 & -325.674 & $0$ & -325.674 & $0$\\ \hline 
			24 & -351.5055 & $C_4$ & -351.5055 & $C_4$\\ \hline 
			25 & -440.1661 & $0$ & -433.6832 & $C_3$\\ \hline 
			26 & -512.0153 & $0$ & -512.0153 & $0$\\ \hline 
			27 & -549.2657 & $0$ & -549.2657 & $0$\\ \hline 
			28 & -656.3142 & $0$ & -656.3142 & $0$\\ \hline 
			29 & -731.1445 & $0$ & -731.1445 & $0$\\ \hline 
			30 & -783.7615 & $0$ & -780.4043 & $D_3$\\ \hline 
			31 & -898.9203 & $C_5$ & -898.9203 & $C_5$\\ \hline 
			32 & -957.8877 & $0$ & -921.6676 & $C_3$\\ \hline 
			33 & -1140.2 & $0$ & -1088.901 & $C_3$\\ \hline 
			34 & -1275.227 & $0$ & -1275.227 & $0$\\ \hline 
			35 & -1386.337 & $0$ & -1386.337 & $0$\\ \hline 
			36 & -1546.171 & $0$ & -1546.171 & $0$\\ \hline 
			37 & -1733.916 & $0$ & -1733.916 & $0$\\ \hline 
			38 & -1909.507 & $S_{12}$ & -1909.507 & $S_{12}$\\ \hline 
			39 & -2033.533 & $0$ & -2033.533 & $0$\\ \hline 
			40 & -2240.636 & $C_3$ & -2240.636 & $C_3$\\ \hline 
			41 & -2488.956 & $0$ & -2488.956 & $0$\\ \hline 
		\end{tabular}
	\end{subtable}
	\begin{subtable}[t]{0.5\linewidth}
		\begin{tabular}[t]{c|cccc} 
			$N$ & $E_{GS}$ & PG$_{GS}$ & $E_{sym}$ & PG$_{sym}$ \\ \hline\hline 
			42 & -2586.568 & $0$ & -2547.455 & $C_2$\\ \hline 
			43 & -2857.83 & $0$ & -2857.83 & $0$\\ \hline 
			44 & -3073.318 & $C_2$ & -3073.318 & $C_2$\\ \hline 
			45 & -3331.461 & $0$ & -3331.461 & $0$\\ \hline 
			46 & -3665.802 & $0$ & -3665.802 & $0$\\ \hline 
			47 & -3903.933 & $0$ & -3903.933 & $0$\\ \hline 
			48 & -4180.57 & $C_4$ & -4107.373 & $D_4$\\ \hline 
			49 & -4631.788 & $0$ & -4512.917 & $C_2$\\ \hline 
			50 & -4746.23 & $C_2$ & -4736.893 & $C_6$\\ \hline 
			51 & -5284.134 & $0$ & -5284.134 & $0$\\ \hline 
			52 & -5684.646 & $0$ & -5653.811 & $C_3$\\ \hline 
			53 & -6128.256 & $0$ & -6128.256 & $0$\\ \hline 
			54 & -6538.604 & $C_2$ & -6538.604 & $S_4$\\ \hline 
			55 & -6800.163 & $0$ & -6800.163 & $0$\\ \hline 
			56 & -7245.993 & $0$ & -7187.049 & $C_2$\\ \hline 
			57 & -7657.339 & $0$ & -7657.339 & $0$\\ \hline 
			58 & -8128.102 & $0$ & -7848.924 & $C_2$\\ \hline 
			59 & -8694.171 & $0$ & -8694.171 & $0$\\ \hline 
			60 & -9022.76 & $0$ & -9022.76 & $0$\\ \hline 
			61 & -9632.079 & $0$ & -9603.014 & $C_2$\\ \hline 
			62 & -10267.12 & $C_2$ & -10267.12 & $C_2$\\ \hline 
			63 & -10652.79 & $0$ & -10597.55 & $C_2$\\ \hline 
			64 & -11566.38 & $0$ & -11566.38 & $0$\\ \hline 
			65 & -12224.11 & $0$ & -12042.76 & $C_2$\\ \hline 
			66 & -12494.04 & $0$ & -12373.88 & $C_3$\\ \hline 
			67 & -13538.89 & $0$ & -12915.27 & $C_2$\\ \hline 
			68 & -14224.9 & $0$ & -14224.9 & $0$\\ \hline 
			69 & -14688.46 & $0$ & -14688.46 & $0$\\ \hline 
			70 & -15824.37 & $0$ & -15824.37 & $0$\\ \hline 
			71 & -16407.1 & $0$ & -16407.1 & $0$\\ \hline 
			72 & -16933.5 & $0$ & -16785.31 & $C_3$\\ \hline 
		\end{tabular}
	\end{subtable}
	\caption{Ground state and highest symmetry solutions on the Tammes lattices.}
	\label{tab:Tammmes}
\end{table*}